\documentclass[sigconf]{aamas} 

\usepackage{balance} 
\usepackage{soul} 

\renewcommand\footnotetextcopyrightpermission[1]{} 
\acmSubmissionID{<<submission id>>}

\title{A Multi-Level Agent-Based Architecture for Climate Governance Integrating Cognitive and Institutional Dynamics}

\author{Ivan Puga-Gonzalez}
\affiliation{
  \institution{NORCE Research AS}
  \city{Kristiansand}
  \country{Norway}
}
\email{ivpu@norceresearch.no}

\author{Önder Gürcan}
\affiliation{
  \institution{NORCE Research AS}
  \city{Kristiansand}
  \country{Norway}
}
\email{ongu@norceresearch.no}

\author{Vanja Falck}
\affiliation{
  \institution{NORCE Research AS}
  \city{Kristiansand}
  \country{Norway}
}
\email{vafa@norceresearch.no}

\author{Christopher Frantz}
\affiliation{
  \institution{NTNU}
  \city{Gjøvik}
  \country{Norway}
}
\email{christopher.frantz@ntnu.no}

\author{F. LeRon Shults}
\affiliation{
  \institution{NORCE Research AS}
  \city{Kristiansand}
  \country{Norway}
}
\email{lesh@norceresearch.no}

\author{David Herbert}
\affiliation{
  \institution{University of Bergen}
  \city{Bergen}
  \country{Norway}
}
\email{david.herbert@uib.no}

\author{Larissa Lopes Lima}
\affiliation{
  \institution{NORCE Research AS}
  \city{Kristiansand}
  \country{Norway}
}
\email{lali@norceresearch.no}

\author{Markus Grendstad Rousseau}
\affiliation{
  \institution{NORCE Research AS}
  \city{Kristiansand}
  \country{Norway}
}
\email{maro@norceresearch.no}

\begin{abstract}
Climate governance processes involve complex interactions between heterogeneous citizens, advocacy groups, media actors, and political decision-makers. While agent-based models (ABMs) have been widely used to study environmental policy and socio-ecological systems, many existing approaches focus either on institutional dynamics or individual behavioural mechanisms in isolation. This paper presents a modular multi-level agent-based architecture that integrates empirically grounded cognitive decision models with strategic institutional behaviour within a unified simulation framework. 
The architecture combines (i) motive-based individual decision-making operationalised through the HUMAT and MOA frameworks, (ii) socially embedded influence processes via demographic homophily networks, and (iii) institutional strategy modules for environmental non-governmental organisations (NGOs), media agents, and politicians. Political decisions emerge from the aggregation of multiple signals, including expert input, public mobilisation, party alignment, and media framing. The model is designed to be empirically calibrated through synthetic populations derived from survey data and and institutional parameters informed through Living Lab stakeholder engagement, and to support scenario-based exploration of climate-relevant land-use governance processes.
Rather than presenting empirical results, this paper focuses on the architectural design principles, modular structure, and integration logic of the model. We discuss how this multi-layered approach contributes to the modelling of democratic climate governance and outline pathways for generalization and future validation.
\end{abstract}

\keywords{Agent-Based Modeling, Climate Governance, Multi-Level Architecture, Behavioral Decision Models, Institutional Strategy}

\newcommand{\BibTeX}{\rm B\kern-.05em{\sc i\kern-.025em b}\kern-.08em\TeX}

\begin{document}


\pagestyle{fancy}
\fancyhead{}


\maketitle 

\section{Introduction}

Climate governance, particularly at the local and regional levels, involves complex interactions among heterogeneous actors with diverging goals, values, and strategic incentives. Decisions regarding land use, infrastructure development, and environmental protection are shaped not only by formal institutional procedures but also by citizen mobilisation, advocacy campaigns, media framing, and political competition. These processes unfold dynamically over time and are often characterized by feedback loops, contested legitimacy, and shifting public opinion. Recent research on social tipping and climate transitions emphasizes the importance of understanding how collective behavioural change and institutional dynamics co-evolve in such systems \cite{Frontiers2022}. 

A motivating example is democratic urban climate governance in Vestland county, Norway, where infrastructure and land-use proposals can create tensions between environmental protection, economic development, mobility needs, and local political priorities. These cases involve citizens, environmental NGOs, media actors, planners, and elected representatives, whose interactions can shift the political fate of a proposal over time. This provides a practical setting for exploring how behavioural, social, and institutional dynamics jointly shape climate-relevant governance outcomes.

Agent-based modelling (ABM) provides a suitable framework for studying such systems, as it enables the representation of heterogeneous actors, bounded rationality, and decentralised interactions within evolving socio-political environments \cite{Bonabeau2002}. 
Prior work has applied ABMs to environmental governance, climate adaptation, and socio-ecological transitions \cite{Patt2005, Balbi2010, Arneth2014, Hoekstra2017, Beckage2018, Bury2019, Tolk2023,Gurcan2025SSCProClimate}. However, existing models often emphasise either institutional rule structures or individual behavioural mechanisms, with comparatively fewer approaches integrating empirically grounded cognitive decision models and strategic institutional behaviour within a unified architectural framework.

In this paper, we present a modular multi-level agent-based architecture for modelling democratic climate governance processes centred on a land-use proposal lifecycle. The architecture integrates three core layers: (1) a micro-level behavioural layer in which citizens make decisions based on motive importance weights and opportunity–ability constraints; (2) a meso-level social influence layer capturing network-based conformity and information diffusion; and (3) a macro-level institutional layer modelling strategic environmental non-governmental organisations (NGOs), media agents, and political decision-makers. Political outcomes emerge from the structured aggregation of multiple signals, including expert assessments, public mobilisation, party alignment, and media framing.

The primary contribution of this paper lies in its design and integration logic rather than in the presentation of empirical findings. We describe how behavioural theory is operationalized computationally, how institutional strategies are represented, and how these components interact through a proposal lifecycle governing land-use decisions. The architecture is designed to be empirically grounded through synthetic populations derived from survey data while remaining modular and portable across cases. In its current instantiation, the model has been developed and refined in collaboration with stakeholders from the Bergen Living Lab within the Horizon Europe project \textit{PRO-CLIMATE}. By articulating the design principles and structural components of this framework, we aim to contribute to the development of more integrated agent-based approaches to modelling climate governance.


\section{Related Work}

\subsection{ABMs of Climate and Environmental Governance}

As noted above, agent-based modelling has been widely applied to the study of environmental governance, land-use dynamics, and climate adaptation processes \cite{Filatova2023}. Such models have been used to explore resource management, policy compliance, energy transitions, and socio-ecological feedbacks \cite{Frontiers2022}. By representing heterogeneous actors and decentralised interactions, ABMs enable the investigation of emergent outcomes arising from local decision-making and institutional constraints.

In climate governance research, modelling efforts address both institutional policy processes and collective behavioural dynamics, often treating these dimensions separately \cite{Frontiers2022, Filatova2023}. Some approaches focus on rule-based interactions among institutional actors and regulatory mechanisms, while others centre on collective action dynamics, norm diffusion, and transition processes in socio-technical systems. Research on social tipping and climate transitions further highlights the importance of understanding how citizen mobilisation, institutional responses, and normative shifts interact over time \cite{Frontiers2022}. More broadly, recent socio-ecological modelling work emphasises the need to connect behavioural microfoundations with governance and policy-level dynamics within cross-scale frameworks \cite{Filatova2023}.

\subsection{Behavioural Microfoundations in ABMs}

A parallel stream of research emphasizes behavioural realism and cognitive microfoundations within agent-based systems \cite{Farmer2025,Lin2025,Giardini2024}. 
Such models incorporate bounded rationality, value-based preferences, social norms, learning processes, and network-mediated influence. 
By modelling individual cognition and motivation explicitly, these approaches aim to capture heterogeneity in preferences and decision strategies and to explain how individual-level dynamics give rise to collective patterns.

While such models advance the representation of behavioural processes, institutional actors are often simplified or treated as exogenous constraints rather than as strategic agents embedded in political decision processes. This separation between behavioural microfoundations and institutional strategy can limit the capacity of models to explore the co-evolution of public mobilisation and formal governance outcomes.

\subsection{Integrating Behavioural and Institutional Dynamics}

Recent work in socio-ecological modelling increasingly emphasises the need to link behavioural microfoundations with institutional and policy-level dynamics \cite{Filatova2023, Frontiers2022}. Large-scale modelling efforts further stress the importance of improved cross-scale coupling between individual decisions, collective processes, and macro-level outcomes \cite{PNAS2025}. At the same time, methodological analyses of socio-ecological ABMs identify the integration of human decision processes with governance structures as a central and ongoing challenge \cite{Filatova2023}.

In particular, recent analyses of socio-ecological ABMs identify persistent challenges in integrating empirically grounded human decision processes with governance and institutional dynamics while maintaining transparency, modularity, and cross-scale coherence \cite{Filatova2023}. Addressing these challenges requires architectural approaches that explicitly structure interactions across micro-, meso-, and macro-level components while maintaining empirical interpretability.

The architecture presented in this paper is designed in response to these integrative challenges. By combining motive-based decision mechanisms, network-mediated influence processes, and structured political signal aggregation within a unified simulation environment, the model seeks to bridge behavioural and institutional perspectives on democratic climate governance.


\section{Architectural Overview}
\begin{figure}
    \centering
    \includegraphics[width=1\linewidth]{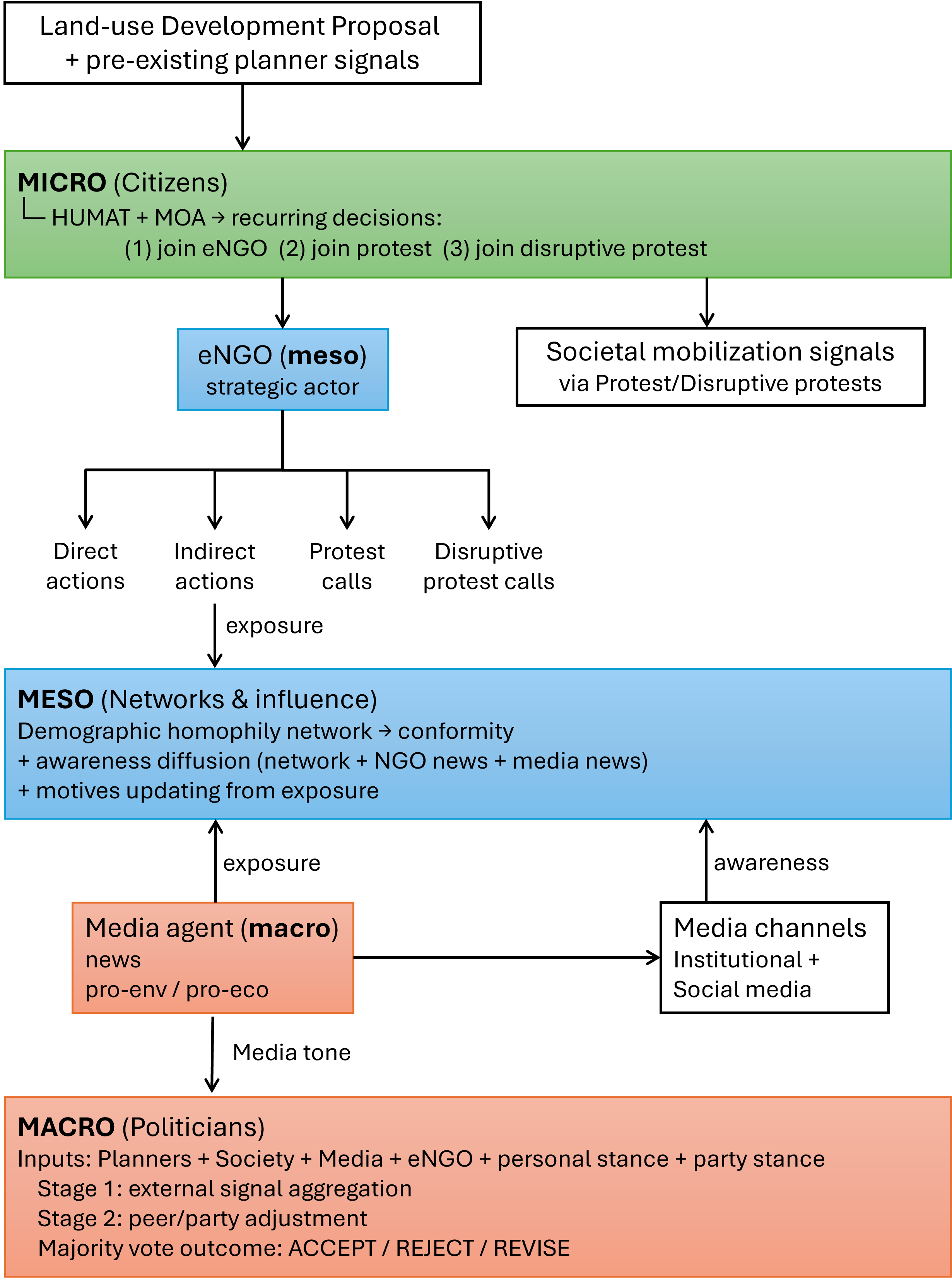}
    \caption{Multi-level agent-based architecture for democratic climate governance.}
    \label{fig:architecture}
\end{figure}

The purpose of the architecture is to model democratic climate governance as a multi-layered system composed of interacting behavioural, social, and institutional subsystems. The overall structure of the architecture and the information flows across levels are illustrated in Figure~\ref{fig:architecture}. The central interaction artefact in the model is a land-use development proposal. The proposal is introduced during initialisation and is loaded together with pre-existing institutional evaluation signals (e.g., from urban planners and other designated expert actors), which serve as structured inputs to the subsequent political aggregation process. Actors embedded in the system interact indirectly through this shared artefact and through network-mediated influence processes.

At a high level, the architecture consists of four agent types: citizens, environmental Non-Governmental Organizations (eNGOs), media agents, and politicians. Citizens represent heterogeneous members of the public whose decisions are shaped by motive importance weights and contextual opportunity–ability constraints. eNGOs act as strategic collective actors capable of mobilizing support or opposition to development proposals. Media agents generate and disseminate information signals that influence public perception. Politicians aggregate multiple inputs—including expert assessments, public mobilisation, party alignment, and media tone—into formal decisions to accept, reject, or revise proposals.

These agents operate across three interconnected layers:

\begin{itemize}
    \item \textbf{Micro-level behavioural layer:} Individual citizens evaluate behavioral options based on motive satisfaction functions and contextual constraints using HUMAT/MOA-derived mechanisms (see Section~\ref{sec:decision-modules}).
    \item \textbf{Meso-level social influence layer:} Citizens are embedded in demographic homophily networks through which conformity and information diffusion processes operate.
    \item \textbf{Macro-level institutional layer:} eNGOs and media agents operate through structured processes that generate information and mobilisation signals feeding into the political aggregation mechanism.
\end{itemize}

As shown in Figure~\ref{fig:architecture}, interactions occur both vertically (micro–meso–macro coupling) and horizontally (e.g., media and eNGO signal generation), with the land-use proposal acting as the central interaction artefact.

\subsection{Design Principles}

The architecture is guided by four explicit design principles.

\textbf{Multi-level integration.} The model connects individual cognitive mechanisms, network-mediated social influence, and institutional dynamics within a unified framework, enabling the study of how public opinion, organized advocacy, media signals, and political aggregation interact.

\textbf{Empirical grounding.} Citizen heterogeneity is initialized using synthetic populations derived from survey data, enabling motive importance weights, opportunity--ability constraints, and media exposure profiles to reflect empirical distributions.

\textbf{Modularity and transparency.} Behavioural, social, and institutional components are implemented as separable modules with clearly defined interfaces, supporting extension and substitution of submodels. 

\textbf{Portability across cases.} While the current instantiation is calibrated to a specific empirical setting, its structure is not case-dependent. The architecture is designed to support upscaling and scenario testing without altering the underlying interaction logic.

\subsection{Simulation Flow and Interaction Logic}

After initialisation, the simulation proceeds in monthly time steps. The sequence of processes depends on whether an eNGO is present. If no eNGO exists, citizens may form one; if an eNGO exists, citizens may decide to join it. When present, the eNGO may undertake actions with selection and intensity modulated by its internal attributes (see Section~\ref{sec:engo-layer}); cumulative action intensity contributes to the pressure signal that later enters politician decision-making.

Each time step includes a media phase in which news may be generated and disseminated through conventional and social media. Citizens receive news depending on their individual exposure profiles, and exposure may update motive importance values (e.g., shifting the relative salience of environmental or economic concerns depending on content). 

The simulation typically runs for 12 time steps (one year). After this period, politicians evaluate the proposal through a two-stage process (external input aggregation followed by peer/party adjustment) and produce a collective majority decision (accept, reject, or revise), which formally terminates the simulation cycle.


The following sections describe each layer in greater detail, beginning with the micro-level behavioural decision architecture.


\section{Micro-Level Behavioural Layer: HUMAT--MOA Decision-Making}

Citizen behaviour in the model is driven by an integrated HUMAT--MOA decision framework, designed to represent psychologically plausible choice while accounting for structural constraints and social influence. In the current instantiation, key motive importances and opportunity/ability proxies are parameterised using survey-response items, but the decision logic is model-agnostic and can be re-parameterised for other contexts.

\subsection{HUMAT: Motive-Based Choice}

Following HUMAT \cite{Jager2025HUMAT,Gurcan2024}, each citizen $i$ evaluates three behavioural decision modules—(a) eNGO membership, (b) action protest participation, and (c) disruptive protest participation—by comparing expected total satisfaction for the available options within each decision (act vs.\ not act). For each decision module $k \in \mathcal{K}$, the agent compares the available options $o \in \mathcal{O}_k$ (e.g., act vs.\ not act) by computing their expected total satisfaction. Each agent has a set of motives $m \in \mathcal{M}$, grouped into (i) value motives (climate concern, nature concern, and growth-first orientation), (ii) experiential motives (perceived economic security), and (iii) social motives (conformity). Each motive has an importance weight $w_{i,m}\in[0,1]$ representing how much that motive matters to agent $i$.

For each option $o \in \mathcal{O}_k$, the agent draws or computes a motive-specific satisfaction $s_{i,m}(o)$ and combines these into a total satisfaction score:
\begin{equation}
S_i(o) \;=\; \sum_{m \in \mathcal{M}} w_{i,m}\, \cdot s_{i,m}(o).
\end{equation}
The behavioural choice within decision module $k$ is then defined as:
\begin{equation}
o_{i,k}^\ast \;=\; \arg\max_{o \in \mathcal{O}_k} S_i(o).
\end{equation}

Satisfaction components are static in structure but are modulated by opportunities and abilities. The importance weights may be dynamic (e.g., updated via network feedback or media exposure). When multiple options yield similar total satisfaction, HUMAT allows for additional tie-breaking through dissonance and heuristics (see \cite{Jager2025HUMAT}).

\subsection{MOA: Motivation, Opportunity, and Ability}

To represent behavioural flexibility, HUMAT is extended with MOA \cite{MacInnis1991MOA}. The key idea is that motivation alone may be insufficient for action: opportunities (external enablers) and abilities (internal or socially mediated capabilities) condition whether and how strongly a given decision is experienced as satisfying.

Operationally, MOA enters as a modifier of satisfaction terms for decisions that are theoretically aligned with a given motive. For example, if joining an eNGO is defined as satisfying an environmental value motive, then low opportunity (e.g., limited time availability) or low ability (e.g., low political self-efficacy) can attenuate the realised satisfaction from that decision. In the current parameterisation, opportunity/ability modifiers are applied only when the underlying motive--decision alignment is positive, i.e., they amplify (or fail to amplify) satisfaction rather than creating motivation where none exists.

\subsection{Decision Modules Implemented in the Simulation}
\label{sec:decision-modules}

The behavioural layer implements three decision modules that recur over time and jointly shape mobilisation dynamics:
(i) forming/joining an eNGO,
(ii) participating in an action protest, and
(iii) participating in a disruptive action protest.
All three modules share the same core structure: for each decision module $k$, agents compare the available options $o \in \mathcal{O}_k$ (act vs.\ not act) using the HUMAT satisfaction aggregation, then apply MOA constraints and network-mediated social influence (Section~\ref{sec:social-influence}) to update their stance and behaviour over time.

\subsubsection{Forming and Joining an Environmental NGO}

If no eNGO exists, citizens may initiate one; if an eNGO exists, citizens decide whether to join. The decision is modelled as a binary choice (join vs.\ not join) with motive importance weights mapped from survey-response items in the current instantiation. For each motive, satisfaction for joining and not joining is specified via distributions whose means reflect the expected direction of alignment (e.g., environmental concern increasing satisfaction from joining, perceived economic insecurity decreasing it). The social motive of conformity is computed endogenously from the agent's network: letting $P_i$ denote the proportion of $i$'s peers who have joined, conformity satisfaction is represented as:
\begin{equation}
s_{i,\text{conf}}(\text{join}) = 2P_i-1,\qquad
s_{i,\text{conf}}(\text{not join}) = 1-2P_i.
\end{equation}

Opportunity and ability factors (e.g., time availability, perceived system responsiveness, political self-efficacy, civic engagement) adjust satisfaction for joining when the corresponding value motives are relevant. Additionally, eNGO presence conditions the availability of the ``join'' option (i.e., joining is only possible if an eNGO exists).

\subsubsection{Participating in an Action Protest}

Citizens also decide whether to participate in an action protest. The action module uses the same motive set as the eNGO module but with decision-specific satisfaction distributions, reflecting differences between organisational membership and episodic action. Participation further includes an awareness constraint: if an agent is not aware that an action is taking place, the participation option is unavailable regardless of the agent's motivation. As in the eNGO module, conformity depends on the proportion of peers who have participated.

\subsubsection{Participating in a Disruptive Action Protest}

Disruptive protest is modelled as a distinct action type with higher perceived social and personal costs. The module uses the same underlying motives and MOA factors but adds two additional constraints. First, agents must be aware that a disruptive protest is occurring. Second, eligibility is restricted via an activist alignment filter derived from survey items capturing attitudes toward disruptive climate activism. In the current instantiation, respondents are classified into attitude categories based on support and disapproval scores, and only the most aligned category is eligible to receive MOA-based satisfaction modifiers for disruptive participation. This mechanism represents the idea that disruptive tactics are normatively acceptable and motivationally viable only for a subset of citizens.

\subsection{Summary}

Together, these HUMAT--MOA modules provide a micro-level account of how heterogeneous citizens transition from inaction to organised membership and various forms of mobilisation under changing informational and social conditions. The next section describes how network-mediated social influence and information diffusion are implemented to connect individual decisions to emergent collective dynamics.

\section{Meso-Level Social Influence Layer}
\label{sec:social-influence}

While the micro-level HUMAT--MOA framework specifies how individual citizens evaluate behavioural options, mobilisation dynamics depend on how citizens are embedded within social networks. The meso-level layer models network-mediated social influence processes that shape conformity, awareness, and (potentially) motive salience over time.

\subsection{Network Structure}

Citizens are embedded in a static social network generated using demographic homophily. Links are formed probabilistically based on similarity across four socio-demographic attributes—age, gender, education, and political orientation. This results in clustered structures that approximate empirically observed patterns of social association. The resulting network is represented as a graph $G = (V, E)$, where $V$ denotes the set of citizen agents and $E$ the set of undirected social ties.

Let $\mathcal{N}_i$ denote the set of neighbours of agent $i$.

\subsection{Conformity and Peer Exposure}

The social motive of conformity is computed dynamically from the observed behaviour of neighbours. For a given decision module $k$, let $P_{i,k}$ denote the proportion of $i$'s neighbours who have chosen the active option (e.g., joining an eNGO or participating in a protest):

\begin{equation}
P_{i,k} = \frac{1}{|\mathcal{N}_i|} \sum_{j \in \mathcal{N}_i} \mathbb{1}(o_{j,k}^\ast = \text{act}),
\end{equation}

where $\mathbb{1}(\cdot)$ is the indicator function.

This proportion enters directly into the conformity satisfaction term defined in Section~\ref{sec:decision-modules}. When peer participation increases, the social satisfaction associated with acting increases correspondingly. In this way, local clustering can generate reinforcement dynamics and collective mobilisation cascades.

\subsection{Awareness Diffusion}

Participation in protest-based decisions requires awareness of the event. Awareness spreads through two channels:

\begin{itemize}
    \item \textbf{Network transmission:} Agents may become aware of an action if they are members of the eNGO calling for participation.
    \item \textbf{Media exposure:} Awareness may also arise from exposure to media signals (see Section~\ref{sec:media-layer}).
\end{itemize}

If an agent is unaware of a protest event at time $t$, the active participation option is unavailable in that time step. Awareness therefore acts as a gating condition in the HUMAT evaluation process.

\subsection{Dynamic Updating of Motive Salience} 

In addition to behavioural diffusion, the model allows for updating of motive importance weights in response to information signals generated by media agents or eNGOs, conditional on individual exposure. For example, repeated exposure to environmental framing may increase the relative salience of environmental value motives. Importantly, the model assumes that information exposure does not decrease motive importance; it can only increase it. 

Conditional on exposure, motive importance is updated using an asymptotic logistic sensitivity function. For a motive $m$ with current importance $w_{i,m}(t) \in [0,1]$, the update rule is:

\begin{equation}
w_{i,m}(t+1) = w_{i,m}(t) + \big(1 - w_{i,m}(t)\big)\,\sigma\big(w_{i,m}(t)\big),
\end{equation}

where the sensitivity term is defined as:

\begin{equation}
\sigma(w) = \frac{1}{1 + e^{-k\,(w - \tau)}},
\end{equation}

with $k$ controlling the steepness of the response curve and $\tau$ defining the importance level at which responsiveness is highest.

This functional form ensures three desirable behavioural properties. First, responsiveness is low when motive importance is very low, reflecting limited receptivity. Second, responsiveness peaks at moderate importance levels. Third, when importance is already high, updating slows asymptotically, preventing overshooting and modelling saturation effects.

Through repeated exposure over time, information signals can therefore amplify existing concerns in a nonlinear and bounded manner.

\subsection{Summary}

The meso-level layer operationalizes how local interaction structures transform individual decision rules into emergent collective mobilisation patterns. Conformity effects, awareness diffusion, and motive updating mechanisms jointly enable nonlinear participation dynamics, including threshold effects and clustered cascades. The next section describes how institutional actors operate at the macro level and how political decisions aggregate signals from citizens, experts, and media.
\section{Macro-Level Institutional Layer}
\subsection{Environmental NGOs: Strategic Mobilisation Agents}
\label{sec:engo-layer}

Environmental NGOs (eNGOs) introduce organised advocacy and strategic pressure into the governance process. In the model, eNGOs are autonomous agents capable of selecting among alternative action repertoires, generating public signals, and exerting pressure on political decision-makers. Their behaviour is structured but heterogeneous, reflecting differences in organisational capacity and strategic orientation.

\subsubsection{Action Repertoire}

Drawing on comparative research on environmental advocacy, the full range of eNGO activities is operationalised into four action categories:

\begin{itemize}
    \item \textbf{Direct Actions:} Institutional engagement strategies such as lobbying, litigation, and formal consultation.
    \item \textbf{Indirect Actions:} Public awareness campaigns and media-based outreach.
    \item \textbf{Protest Actions:} Conventional demonstrations and visible public mobilisation.
    \item \textbf{Disruptive Protest Actions:} High-impact, norm-challenging tactics such as blockades or symbolic disruption.
\end{itemize}

Each category represents a cluster of empirically observed strategies, simplified into algorithmically tractable action types.

\subsubsection{Organisational Attributes}

Each eNGO is characterised by five internal attributes:

\begin{itemize}
    \item \textbf{Resources} $R \in [0,1]$: Financial and organisational capacity.
    \item \textbf{Experience} $E \geq 0$: Organisational maturity (years active).
    \item \textbf{Strategic Orientation} $S \in [0,1]$: Degree of confrontational vs.\ collaborative stance.
    \item \textbf{Cultural Trust} $T \in [0,1]$: Confidence in institutional responsiveness and legitimacy.
    \item \textbf{Political System Openness}: Treated as open in the current implementation.
\end{itemize}

These attributes condition which actions are eligible and how strongly they influence the system.

\subsubsection{Action Eligibility and Selection}

Each action type is subject to eligibility conditions based on organisational attributes. For example:

\begin{itemize}
    \item \textbf{Direct Actions} require high resources and institutional trust.
    \item \textbf{Protest Actions} require moderate resources and are more likely when strategic orientation is confrontational.
    \item \textbf{Disruptive Actions} require a strongly confrontational orientation.
    \item \textbf{Indirect Actions} require only minimal resources for dissemination via social media, but higher resource levels for access to conventional media channels (e.g., television, radio, newspapers).
\end{itemize}

Eligibility does not imply execution. At each time step, eligible actions are selected probabilistically, with probabilities scaled by relevant organisational attributes (e.g., confrontational orientation increases likelihood of protest selection).

\subsubsection{Action Intensity}

When an action is executed, it generates an intensity score reflecting its influence on the political environment. Intensity is computed as a weighted function of organisational attributes and mobilisation size (for protest and disruptive protest actions).

For institutional (direct) actions:
\begin{equation}
I_{\text{direct}} = 10R + 8E + 5T.
\end{equation}

For protest-based actions:
\begin{equation}
I_{\text{protest}} = 5R + 5E + 10S_p,
\end{equation}
where $S_p$ denotes protest size (proportion of participating citizens).

These formulations reflect the assumption that institutional capacity drives direct engagement effectiveness, while visible participation size drives protest impact.

\subsubsection{Cumulative Pressure and Diminishing Returns}

Action effects accumulate over time but exhibit diminishing marginal influence. Let $I_t$ denote the intensity of an action at time $t$. Cumulative pressure $C_t$ evolves according to a nonlinear accumulation function:

\begin{equation}
C_{t+1} = C_t + \gamma I_t \exp(-\delta C_t),
\end{equation}

where $\gamma$ controls scaling and $\delta$ governs diminishing returns.

This mechanism captures saturation effects: repeated lobbying or repeated protest actions yields progressively smaller marginal impact. The cumulative pressure variable becomes one of the structured inputs into political decision-making (Section~\ref{sec:political-layer}).

\subsubsection{Indirect Actions as Information Signals}

Indirect actions differ structurally from institutional and protest-based actions. Rather than generating direct political pressure, they produce information signals that diffuse through media channels and affect citizen-level motive salience.

When an indirect action is selected, the eNGO generates one information item. Dissemination channels depend on organisational resources:

\begin{itemize}
    \item \textbf{Social Media Dissemination:} Available to all eNGOs with minimal resources.
    \item \textbf{Conventional Media Dissemination:} Accessible only to eNGOs exceeding a higher resource threshold, with probability of coverage increasing in $R$.
\end{itemize}

Citizen exposure depends on individual media consumption profiles. Agents who are members of the eNGO are assumed to receive the signal deterministically. Non-member citizens receive the signal probabilistically based on their degree of exposure to social or conventional media channels.

Upon exposure, motive importance weights are updated according to the logistic updating rule defined in Section~\ref{sec:social-influence}. Indirect actions therefore operate through a salience-amplification mechanism rather than through cumulative pressure. Their systemic effect is mediated via citizen behaviour, which may subsequently influence protest size, conformity dynamics, and ultimately political aggregation.

\subsubsection{Systemic Role}

Through repeated action selection, intensity generation, and cumulative pressure accumulation, eNGOs introduce organised strategic behaviour into the governance system. They influence citizens indirectly through mobilisation and information signalling, and influence politicians directly via aggregated pressure signals.
\subsection{Media Agents: Information Framing and Signal Generation}
\label{sec:media-layer}

In addition to organised advocacy by eNGOs, the model includes a media agent representing institutional news outlets (e.g., newspapers, television, radio). The media agent operates continuously and shapes the informational environment through stochastic news generation and multi-channel dissemination. Unlike eNGOs, whose actions are strategic and proposal-focused, the media agent provides background framing signals that modulate motive salience across the population.

\subsubsection{News Generation Process}

At each time step, the media agent generates a news item with probability $\rho_f$ (news frequency). When a news item is generated, its directional framing is determined stochastically:

\begin{itemize}
    \item With probability $\rho_e$, the news is framed as pro-environment.
    \item With probability $1 - \rho_e$, the news is framed as pro-economic.
\end{itemize}

Each news item is disseminated via institutional media by default. With probability $\rho_s$, the same item is also shared through social media channels. This structure reflects hybrid media ecosystems in which mainstream content may be amplified through digital platforms.

\subsubsection{Directional Framing Effects}

News framing affects which motives are amplified at the citizen level:

\begin{itemize}
    \item Pro-environment news increases the importance of environmental value motives (climate and nature concern).
    \item Pro-economic news increases the importance of experiential (economic security) and value motives (growth-first orientation).
\end{itemize}

Upon exposure, motive importance weights are updated according to the logistic updating rule defined in Section~\ref{sec:social-influence}.

\subsubsection{Heterogeneous Exposure Probabilities}

As for the eNGO information signals, citizens differ in their likelihood of encountering information signals from media agents. Each agent is assigned exposure probabilities for institutional media and social media based on self-reported media consumption profiles in the empirical data used for parameterisation.

Institutional media exposure probability increases with the number of traditional news sources followed (television, radio, newspapers). Social media exposure probability increases with reported use of digital platforms and online sources. Agents with no reported media engagement have a minimal baseline probability of exposure.

Formally, for a given channel $c \in \{\text{institutional}, \text{social}\}$, exposure at time $t$ is modelled as a Bernoulli trial with agent-specific probability $p_{i,c}$. Only agents exposed to a given news item update their motive importance weights.

\subsubsection{Systemic Role}

The media agent operates as an exogenous but persistent framing mechanism within the governance system. By modulating motive salience across the population, media signals influence mobilisation decisions indirectly through the HUMAT--MOA framework, and enter political aggregation through the media-tone signal and its salience weighting rather than through a dedicated cumulative-pressure stock variable.

\subsection{Political Aggregation Mechanism}
\label{sec:political-layer}

Politicians represent the final decision-makers in the governance system. They evaluate the development proposal by aggregating technical assessments, public signals, organised advocacy, media framing, and internal political alignment. Rather than acting as strict optimisers, politicians are modelled as boundedly rational agents who balance electoral incentives, ideological commitments, and organisational cohesion.

\subsubsection{Two-Stage Decision Process}

Political decision-making unfolds in two sequential stages:

\paragraph{Stage 1: External Signal Aggregation}

Each politician integrates six inputs:

\begin{itemize}
    \item \textbf{Institutional Assessment} ($U$): Technical evaluations from planners and formal institutions.
    \item \textbf{Societal Signal} ($S$): Public mobilisation level.
    \item \textbf{Media Tone} ($M$): Aggregate directional framing of coverage.
    \item \textbf{eNGO Pressure} ($D$): Cumulative strategic pressure from organised advocacy.
    \item \textbf{Personal Stance} ($P$): The politician’s predefined ideological orientation.
    \item \textbf{Party Stance} ($PP$): Average ideological position of their party.
\end{itemize}

Inputs are first transformed to reflect direction and intensity. Institutional signals (U) are centered around neutrality and modulated by framing effects. Societal and media signals (S, M) are log-ratio transformed and normalized via a $\tanh$ function to capture nonlinear majority effects. Directional inputs ($D$, $P$, $PP$) are split into acceptance and rejection components.

\paragraph{Weight Modulation}

Each input has a baseline weight. Weights are adjusted by:

\begin{itemize}
    \item \textbf{Salience scaling} (for society and media), reflecting issue visibility.
    \item \textbf{Strategic orientation multipliers}, determined by whether the politician is vote-seeking, policy-seeking, or office-seeking.
\end{itemize}

Adjusted weights are normalized to sum to one before aggregation.

\paragraph{Utility Calculation}

Two utilities are computed:

\begin{equation}
U_{\text{accept}} = \sum_i W_i \cdot X_{i,\text{accept}},
\end{equation}

\begin{equation}
U_{\text{reject}} = \sum_i W_i \cdot X_{i,\text{reject}}.
\end{equation}

A decisiveness threshold $\delta$ determines the preliminary decision:

\begin{itemize}
    \item Accept if $U_{\text{accept}} > U_{\text{reject}} + \delta$,
    \item Reject if $U_{\text{reject}} > U_{\text{accept}} + \delta$,
    \item Otherwise request revision.
\end{itemize}

\paragraph{Stage 2: Peer and Party Adjustment}

After computing a preliminary decision, politicians may revise their stance through peer comparison. Each politician evaluates the weighted support for each decision option among peers, where influence is proportional to party similarity.

For politician $i$, support for decision $d$ is:

\begin{equation}
\text{Support}_d^i = \sum_{j \neq i,\, D_j = d} \text{Sim}(P_i,P_j),
\end{equation}

where similarity equals 1 for same-party peers and decreases with ideological distance otherwise.
The probability of maintaining or switching position is proportional to peer support levels. This mechanism captures intra-party negotiation and cohesion dynamics while preserving stochastic heterogeneity.

\subsubsection{Collective Outcome}

After Stage 2 adjustments, the final outcome is determined by majority vote among politicians. The option receiving the highest number of votes (Accept, Reject, or Revision) becomes the collective decision, concluding the simulation cycle.


\section{Empirical Instantiation, Calibration and Validation Strategy}

\subsection{Case Context and Model Instantiation}

To demonstrate empirical grounding and practical applicability, the architecture has been instantiated in the context of democratic urban climate governance in Vestland county (Norway). The case concerns infrastructure and land-use proposals that generate tensions between environmental protection and economic development—an archetypal governance dilemma involving citizens, advocacy organisations, media actors, planners, and elected officials.

The Vestland instantiation serves as a proof of concept for the broader architecture. While the structural design of the model is not case-dependent, parameters and agent attributes are calibrated to reflect the empirical characteristics of this governance setting. This ensures that the model operates within realistic demographic, political, and informational conditions.

\subsection{Calibration Strategy}

Calibration has been conducted across the main subsystems of the model.
The synthetic citizen population was constructed using data from the Norwegian Citizen Panel and the Norwegian Citizen Survey, ensuring that demographic composition, motive importance distributions, and media exposure profiles reflect empirically observed patterns. Politician agents were calibrated to the real composition of the relevant political body, including the number of representatives and their party affiliations. Party stances on the development–sustainability dimension and ideological similarity indices were developed in collaboration with stakeholders in the Bergen Living Lab, including local political and civil-society actors with first-hand knowledge of the governance context. This Living Lab engagement ensured that institutional parameters reflected locally grounded interpretations rather than abstract assumptions.
eNGO attributes (e.g., resources, experience, strategic orientation) were informed by empirical information obtained through stakeholder engagement and case-specific knowledge. Media parameters were calibrated using empirical data on the volume and tone of news items related to the proposal.

This calibration is iterative rather than definitive. Parameter values are treated as empirically informed anchors that will be further refined through ongoing validation and sensitivity analysis.

\subsection{Sensitivity Analysis}

Sensitivity analysis is conducted in two stages.
First, subsystem-level sensitivity tests isolate individual components (eNGOs, politicians, society) by manually varying their parameters while holding others constant. This allows identification of parameter ranges that meaningfully affect decision outcomes and clarifies the relative influence of strategic orientation, mobilisation intensity, and public sentiment.
Second, full-model sensitivity analysis evaluates the integrated system under dynamic interaction. Here, mobilisation, media generation, motive updating, and political aggregation co-evolve endogenously. This stage serves as a robustness assessment, testing the stability of collective outcomes under plausible parameter variation.
Preliminary subsystem-level analyses have been completed for eNGOs and political aggregation, with full-system sensitivity analysis underway.

\subsection{Validation Approach}

Validation combines quantitative comparison and qualitative plausibility assessment. Where empirical data are available, model outputs are compared to observed indicators such as:
\begin{itemize}
    \item patterns of citizen mobilisation,
    \item voting behaviour of political representatives,
    \item media volume and framing.
\end{itemize}

In areas with limited data availability, validation relies on structured stakeholder engagement within the Bergen Living Lab, allowing practitioners to assess behavioural plausibility, institutional realism, and scenario relevance. This hybrid validation strategy reflects the complexity of governance systems, where not all relevant mechanisms are directly observable.
Uncertainties and potential biases—particularly in areas of limited data—are explicitly documented and considered when interpreting model behaviour.

\subsection{Reproducibility and Availability}

The model has been developed within the Horizon Europe project \textit{PRO-CLIMATE} (grant number 101137967). The implementation is publicly available in an open repository\footnote{\url{https://gitlab.norceresearch.no/cmss/digital-social-twins/models/pro-climate}}
 and the full technical documentation is archived on Zenodo\footnote{\url{https://zenodo.org/records/18400568}}
.
The model architecture also builds on the earlier PRO-CLIMATE multi-agent computer model deliverable \cite{PugaGonzales2025PROCLIMATED52}.
Providing access to both code and documentation supports transparency, reproducibility, and future reuse of the architecture in other governance contexts.
\section{Conclusion and Outlook}

This paper presented a multi-level agent-based architecture for modelling democratic climate governance, integrating cognitive decision-making, network-mediated social influence, strategic organisational behaviour, and institutional political aggregation within a unified framework.

At the micro level, citizen behaviour is represented using a HUMAT–MOA decision structure that captures motive heterogeneity, opportunity/ability constraints, and bounded rationality. At the meso level, demographic homophily networks enable conformity dynamics, awareness diffusion, and nonlinear updating of motive salience. At the macro level, strategic eNGO action repertoires, media signal generation, and weighted political aggregation mechanisms translate mobilisation and information signals into formal collective decisions.

The core contribution of this work lies not in a single behavioural mechanism, but in the architectural integration of these components. By combining empirically grounded cognitive modelling with strategic organisational behaviour and institutional decision rules, the framework bridges gaps that often separate behavioural and institutional modelling traditions.

As such, the model contributes to ongoing efforts within the agent-based modelling community to develop architectures capable of representing complex socio-political systems in a transparent, modular, and empirically grounded manner.
Future work may further connect this architecture to LLM-powered social digital-twinning platforms to support interactive scenario exploration, stakeholder-facing experimentation, and more flexible interfaces for climate governance simulations \cite{Gurcan2026PAAMS}.



\begin{acks}
Disclaimer: Funded by the European Union (grant number 101137967). Views and opinions expressed are, however, those of the author(s) only and do not necessarily reflect those of the European Union or the European Climate, Infrastructure and Environment Executive Agency (CINEA). Neither the European Union nor CINEA can be held responsible for them.
\end{acks}



\bibliographystyle{ACM-Reference-Format} 
\bibliography{references}


\end{document}